# Macro-spin Modeling and Experimental Study of Spin-orbit Torque Biased Magnetic Sensors

Yanjun Xu[1,2], Yumeng Yang[1], Ziyan Luo[1], Baoxi Xu[2], and Yihong Wu[1,*]

[1]*Department of Electrical and Computer Engineering, National University of Singapore, 4 Engineering Drive 3, Singapore 117583, Singapore*

[2]*Data Storage Institute, A*STAR (Agency for Science, Technology and Research), 2 Fusionopolis Way, 08-01 Innovis, Singapore 138634, Singapore*

We reported a systematic study of spin-orbit torque biased magnetic sensors based on NiFe/Pt bilayers through both macro-spin modeling and experiments. The simulation results show that it is possible to achieve a linear sensor with a dynamic range of 0.1 - 10 Oe, power consumption of 1μW – 1mW, and sensitivity of 0.1-0.5 Ω/Oe. These characteristics can be controlled by varying the sensor dimension and current density in the Pt layer. The latter is in the range of $1 \times 10^5$ - $10^7$ A/cm$^2$. Experimental results of fabricated sensors with selected sizes agree well with the simulation results. For a Wheatstone bridge sensor comprising of four sensing elements, a sensitivity up to 0.548 Ω/Oe, linearity error below 6%, and detectivity of about 2.8 nT/√Hz were obtained. The simple structure and ultrathin thickness greatly facilitate the integration of these sensors for on-chip applications. As a proof-of-concept experiment, we demonstrate its application in detection of current flowing in an on-chip Cu wire.

[*] Author to whom correspondence should be addressed: elewuyh@nus.edu.sg

## I. INTRODUCTION

Magnetic biasing and stabilizing techniques are commonly used to enhance the linearity, sensitivity and dynamic range of magnetic sensors, and to suppress the Barkhausen noise induced by domain wall motion.[1-4] The purpose of magnetic biasing is to set up a proper magnetic configuration for the sensing layer at zero external field so as to maximize the sensor's dynamic range, sensitivity and linearity, whereas the primary role of stabilizing field is to suppress the domain wall motion. Optimization of the biasing field strength and direction, particularly for sensors for weak field detection, is a delicate task; a weak biasing field will lead to instability and nonlinearity whereas a strong bias field tends to degrade the sensor's sensitivity. In addition to field strength and direction, equally important is the uniformity of the bias field. Take anisotropic magnetoresistance (AMR) sensor as an example, the two most commonly used biasing schemes are soft-adjacent layer (SAL) biasing and barber pole biasing. As shown in Fig.1a, in the case of SAL biasing, a soft magnetic layer, or SAL, is layered with the sensing layer via an insulating spacer. The SAL is typically chosen such that it will have a higher resistivity and lower coercivity than the sensing layer. Therefore, when a current is applied to the trilayer structure, a large portion of it will flow through the sensing layer and thereby generating a magnetic field. The field generated by the sensing current will push the magnetization of the SAL off the easy axis, which in turn will generate a stray field and bias the magnetization of sensing layer away from the current direction. When the thickness and magnetization of both the sensing and SLA layers are optimized, the angle between the current and the magnetization of the sensing layer can be readily set to be $45^o$ by adjusting the sensing current. While the SAL scheme was successfully implemented in AMR sensors, it requires dedicated process work to optimize the structure and, moreover, it also suffers the drawback of non-uniformity in the biasing field, particularly at the edges. On the other hand, in the case of barber pole biasing, instead of changing the magnetization direction, the local current is directed away from the ease axis direction by patterned conducting strips deposited directly atop the sensing layer (Fig.1b). The strips are aligned at an angle of 45° from the ease axis direction of the sensing element. In this way, when a current is supplied from the two end contacts, electrical current between the neighboring strips will form a 45° angle from the magnetization direction and thereby leading to a linear response to transverse field. It is clear from the design that the barber pole design require additional process steps and non-uniformity also exists at the edges.

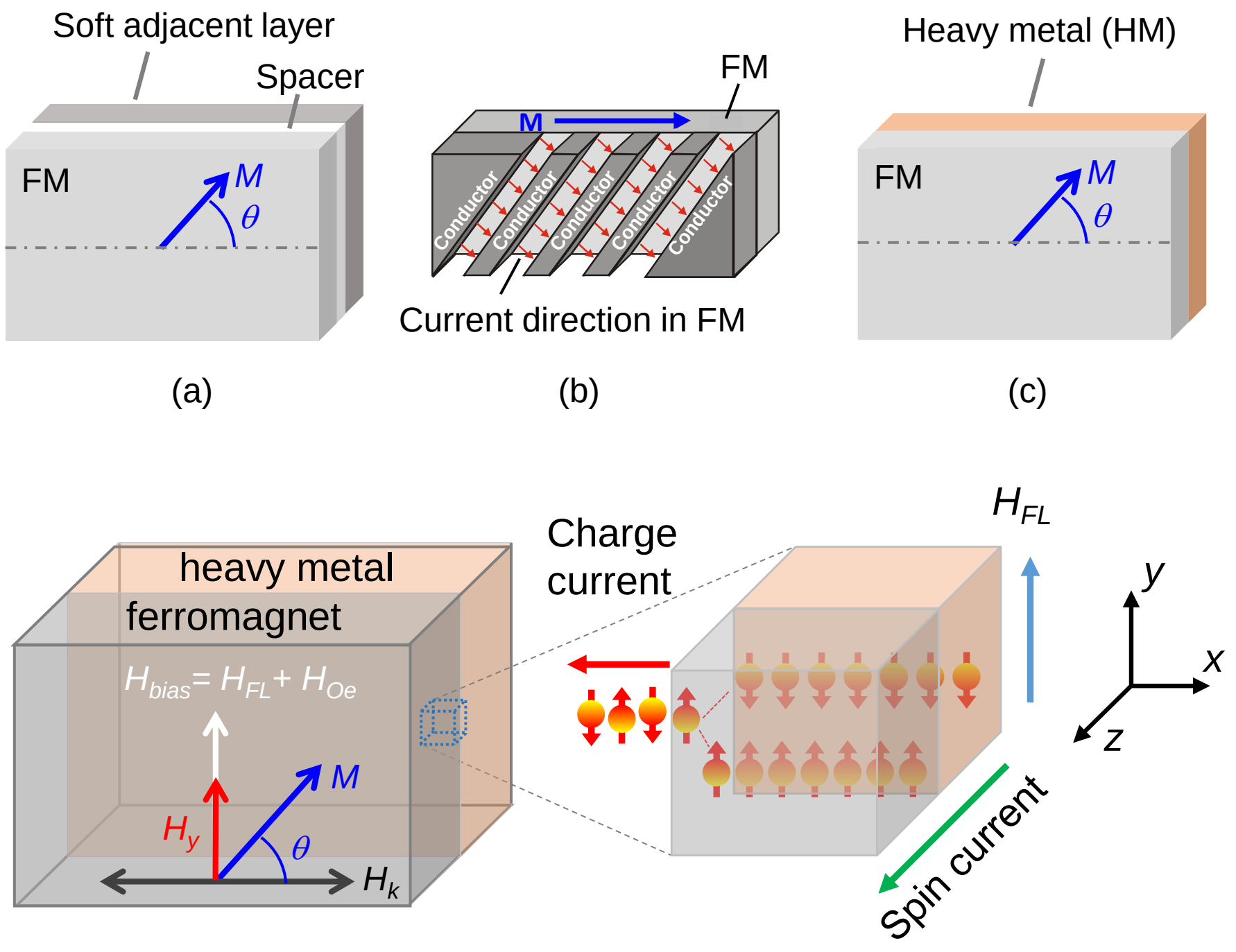

(d)

Fig.1. Different types of transverse bias schemes for AMR sensors: (a) soft-adjacent layer biasing, (b) barber-pole biasing, and (c) spin-orbit torque biasing. (d) illustration of field-like effect field ($H_{FL}$) functioning as a transverse bias together with the Oersted field ($H_{Oe}$) generated by the current in the heavy metal layer.

To simplify the biasing structure and at the same time provide a uniform bias across the active area of the sensor, we have recently introduced a biasing technique based on spin-orbit torque[5] (SOT) (see Fig.1c). The SOT, present in ferromagnet (FM) / heavy metal (HM) heterostructures, has been widely studied as a promising mechanism for switching the magnetization of ultrathin FM layers. There are two types of SOTs, one is called field-like (FL) and the other is called (anti)damping-like (DL). Associated with the FL SOT is an effective field which is transverse to the charge current (Fig.1d). As demonstrated in our recent work, the FL effective field is uniquely suited for transverse magnetic biasing given its high uniformity and good controllability. The use of SOT biasing greatly simplifies the sensor structure and, in fact, what one needs is only a FM/HM bilayer. This made it possible to realize a semi-transparent sensor. In this paper, we report a systematic study of SOT-biased AMR sensors by focusing on how the dimension of the sensor would affect its dynamic range, linearity, sensitivity and power consumption and the current density range that is required to achieve the desirable performance. We will first present the simulation results based on the macro-spin model and then describe the experimental results. The application of SOT-biased sensor in on-chip current detection will also be demonstrated.

## II. MODELING OF SOT BIASED MAGNETORESISTANCE SENSORS

### A. Sensor linearization by field-like SOT effective field

The operation of SOT-biased sensor can be modeled using the macro-spin model by including the FL effective filed. Upon inclusion of $H_{FL}$ in the model, the role of HM can be neglected as it only functions as a current shunting element. The equilibrium direction of the magnetization can be found through minimization of the free energy density $\varepsilon$ of the FM element, which is given by:[6]

$$\varepsilon = \frac{\mu_0}{2} M_s^2 N_x + \frac{\mu_0}{2} M_s\left[M_s\left(N_y - N_x\right) + H_k\right] sin^2\theta - \mu_0 M_s\left(H_{bias} + H_y\right) sin\theta \quad (1)$$

where $H_{bias} = H_{FL} + H_{Oe}$, $N_x$ and $N_y$ are the demagnetizing factors in $x$- and $y$-direction, respectively, $H_k$ is the induced uniaxial anisotropy field, $H_y$ is the external field, and $\theta$ is the angle between the easy axis and the magnetization. The demagnetizing energy due to $M_z$ component is negligible due to in-plane anisotropy of the film. Here, $H_{Oe} = \frac{t_{HM} j_{HM}}{2}$, is the Oersted field generated from the current in the HM layer with $j_{HM}$ the current density and $t_{HM}$ the thickness of HM. As we will discuss in the experimental section, the magnetic sensor fabricated has a flat ellipsoidal shape with a dimension of $a \times b \times t_{FM}$; here, $a > b >> t_{FM}$ and $t_{FM}$ is the thickness of the sensor element. In such case, $N_x$ and $N_y$ can be calculated analytically as follows:[7]

$$N_x = \frac{t_{FM}}{a}\frac{K-E}{e^2}(1-e^2)^{1/2},\ N_y = \frac{t_{FM}}{a}\frac{E-K(1-e^2)}{e^2(1-e^2)^{1/2}} \qquad (2)$$

Here, $K$ and $E$ are complete elliptic integral of the first and second kind, respectively, with the argument $e = (1 - b^2/a^2)^{1/2}$. Minimization of $\varepsilon$ gives $sin\theta = \frac{H_{bias}+H_y}{H_d+H_k}$, where $H_d = M_s(N_y - N_x)$. For FM/HM bilayers, when both FM is thin, in addition to conventional AMR, spin Hall magnetoresistance (SMR) also becomes important. For in-plane film with negligible $M_z$ component, the resistivity of the FM layer can be expressed as $\rho = \rho_0 + (\Delta\rho_{AMR} + \Delta\rho_{SMR})m_x^2$; here $m_x$ is the normalized magnetization component in $x$-direction, which in this case is simply $cos\theta$. Therefore, when the FM/HM bilayer is used as a sensor, the output signal is given by

$$\Delta V = \Delta R \alpha I cos^2\theta \qquad (3)$$

Here, $\Delta R$ is the resistance change when the magnetization is rotated from $y$- to $x$-direction, $I$ is the current applied to the bilayer, and $\alpha < 1$ accounts for the current shunting effect by the HM layer. As we reported previously[5], at a NiFe thickness of 2 nm, the SMR is about two times of AMR, and therefore, the MR is dominantly of SMR characteristic. To obtain a linear response, the magnetization is usually biased at 45° away from the easy axis and at the same time ensure that the external field is much smaller than the bias field. In this case, Eq. (3) can be approximated as $\Delta V = \Delta R \alpha I \left(\frac{1}{2} - \frac{\sqrt{2}H_y}{H_k+H_d}\right)$ when $H_y \ll H_k + H_d$. As we will discuss shortly in the experimental section, $H_{Oe}$ and $H_k$ are typically much smaller than $H_{FL}$ and $H_d$. Therefore, for obtaining a linear response, one requires that $H_{FL} \approx (H_d + H_k)/\sqrt{2}$. The above analysis demonstrated clearly that it is possible to achieve a linear sensor despite the very simple structure. In addition to linearity, for practical applications, one is also concerned about the dynamic range, sensitivity, field resolution, and power consummation. In what follows, we analyze how these characteristics vary with the dimension and material properties of the FM/HM bilayers.

**B. Dynamic range**

The dynamic range of the sensor is determined by $H_d + H_k$. The anisotropy field $H_k$ is strongly dependent on the film thickness, whereas $H_d$ is determined by both the dimension and saturation magnetization of the FM layer. As we will discuss shortly, in order to maximize the SOT effect, the FM layer has to be made as thin as possible. As we reported previously, in the case of NiFe, the smallest thickness is around 1.8 nm, below which the NiFe layer (capped by Pt) behaves like a superparamagnet. At this thickness, the anisotropy field is typically around 0.5 Oe. Once the thickness is fixed, the next critical parameter that affects the demagnetizing field are both the lateral dimension, *a* and *b*, and the aspect ratio, *a/b*. A large aspect ratio is desirable for domain stability, but too large an aspect ratio may result in large power consumption as it requires a larger current to bias the sensor into linear range. Fig.2a shows the calculated $H_d$ as a function of length *a* for *a/b* = 4 and 8, respectively. The thickness of FM layer is fixed at 2 nm. The calculations were performed for different $M_s$, *i.e.*, 400, 600, and 800 emu/cm$^3$, respectively. Note that 800 emu/cm$^3$ corresponds to the $M_s$ of bulk NiFe, but for ultrathin films, the $M_s$ is typically smaller and its exact value may vary from sample to sample. The calculation results in Fig.2a show clearly that the dynamic range of the sensor can be varied in a large range through controlling the sensor size. However, in the present case of SOT-biasing, the actual dynamic range eventually has to be determined by the size of the FL effective field, $H_{FL}$.

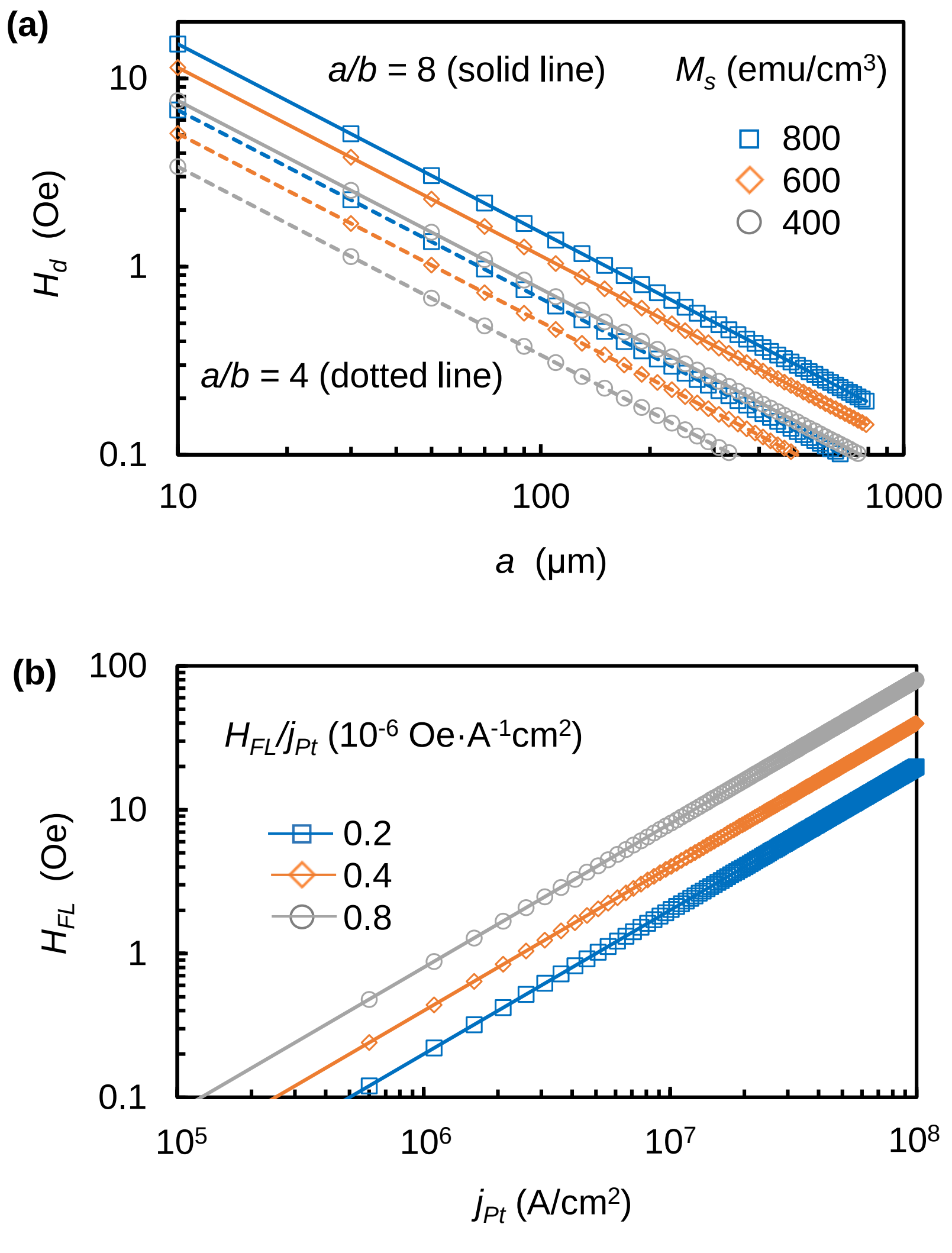

Fig.2. (a) Calculated demagnetizing field as a function of long axis length, $a$, for $a/b$ = 4 (dotted-line) and 8 (solid-line), respectively, and $t_{FM}$ = 2 nm; symbols correspond to different $M_s$ values. (b) Calculated FL effective field in NiFe as a function of current density in Pt with $H_{FL}/j_{Pt}$ = 0.2, 0.4, and 0.8×10$^{-6}$ Oe·(A$^{-1}$cm$^2$), respectively.

As we mentioned in the introduction, both DL and FL torques are present in FM/HM heterostructures. It is generally accepted that both spin Hall effect (SHE)[8-11] and Rashba-Edelstein (RE)[12-15] interaction contribute to generating both types of torques, but their respective roles in the two types of torques are still debatable. The RE interaction is considered to contribute more in generating the FL torque, however, recently there is growing evidence to suggest that SHE also plays an important role in generating the field-like torque.[11,16-20] Therefore, the SOT induced field-like effective field can be estimated from the relation:[19,20]

$$\frac{H_{FL}}{j_{HM}} = \frac{\hbar}{2e}\frac{\theta_{SH}}{\mu_0 M_s t_{FM}}\left[1 - \frac{1}{\cosh(t_{HM}/\lambda_{HM})}\right]\frac{g_i}{(1+g_r)^2+g_i^2} \tag{4}$$

where $g_r = Re[G_{mix}]\rho_{HM}\lambda_{HM}\coth(t_{HM}/\lambda_{HM})$ , $g_i = Im[G_{mix}]\rho_{HM}\lambda_{HM}\coth(t_{HM}/\lambda_{HM})$ with $G_{MIX}$ the spin mixing conductance of FM/HM interface, $\theta_{SH}$ the spin Hall angle of HM, $\rho_{HM}$ the resistivity of HM, $t_{FM}$ ($t_{HM}$) the thickness of FM (HM) and $\lambda_{HM}$ the spin diffusion length in HM, $M_s$ the saturation magnetization of FM layer, $\hbar$ the reduced Planck constant, $\mu_0$ the vacuum permeability, and $e$ the electron charge. The key result of Eq.(4) is that $H_{FL}/j_{HM}$ is inversely proportional to $M_s t_{FM}$. As $M_s$ decreases with $t_{FM}$, a thinner FM will lead to a larger effective field due to reduction of both $M_s$ and $t_{FM}$. Since the spin mixing conductance is sample dependent, Nan *et al.*[16] have introduced an effective spin Hall angle $\theta_{FL}$ for NiFe/Pt bilayer and express the FL effective field to current density ratio as $\frac{H_{FL}}{j_{Pt}} = \frac{\hbar}{2e}\frac{\theta_{FL}}{\mu_0 M_s t_{NiFe}}$. Based on the reported $\theta_{FL}$value of 0.024 and $M_s$ value of 300 – 500 emu/cm$^3$ for ultrathin NiFe, the $H_{FL}/j_{Pt}$ ratio is estimated to be in the range of 0.39 – 0.76×10$^{-6}$ Oe·(A$^{-1}$ cm$^2$). Fig.2b shows the calculated $H_{FL}$ as a function of $j_{Pt}$ by setting $H_{FL}/j_{Pt}$ at 0.2, 0.4, and 0.8×10$^{-6}$ Oe·(A$^{-1}$ cm$^2$), respectively. As can be seen, with a current density of 2×10$^5$ - 3×10$^7$ A/cm$^2$, the effective field can cover the entire range of 0.1 – 10 Oe in

Fig.2a. These results demonstrate clearly that it is possible to use the SOT effective field for sensor linearization.

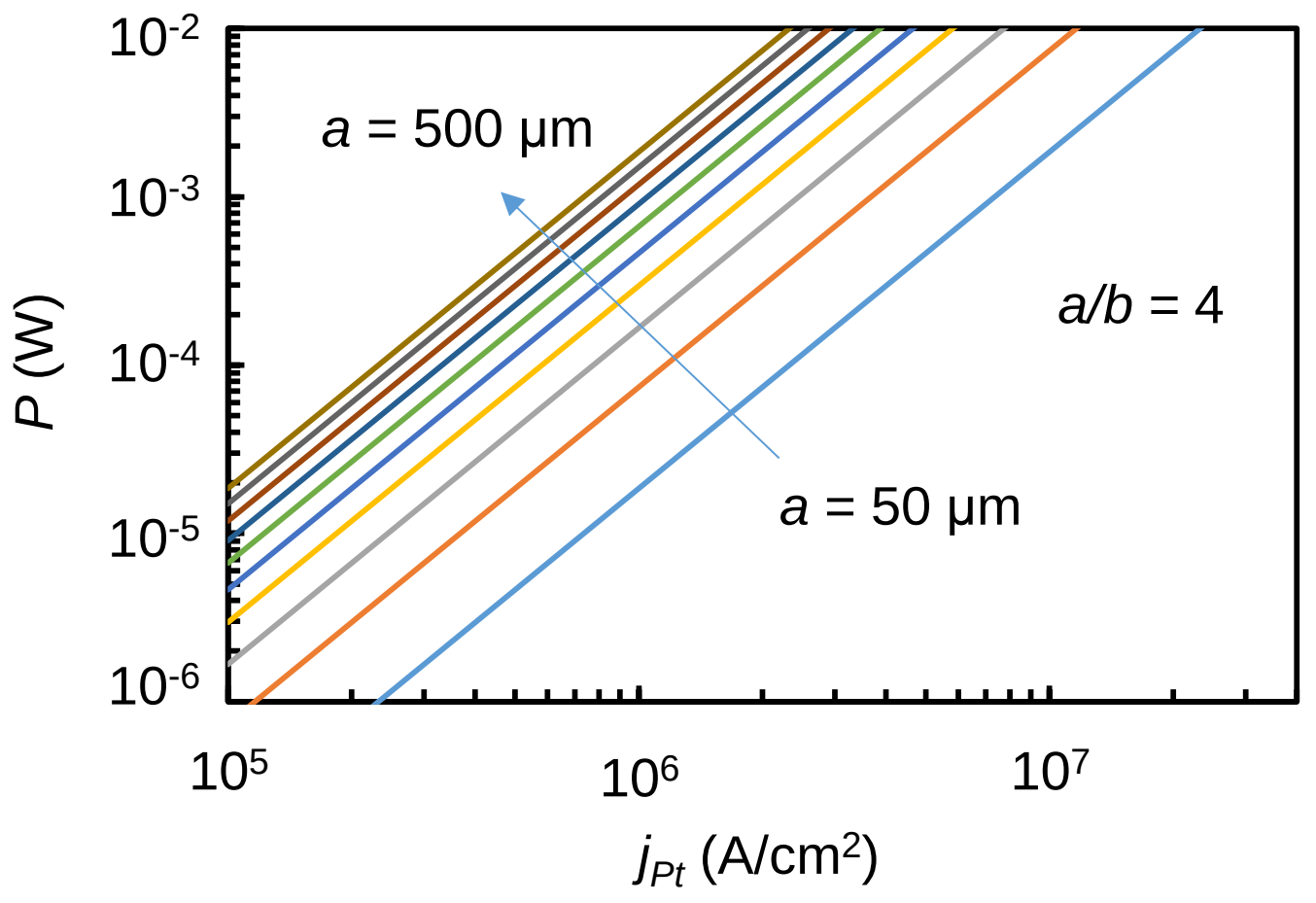

Fig.3. Calculated power consumption as a function of current density in Pt. Parameters used in the calculation are $a/b = 4$, $t_{NiFe} = t_{NiFe} = 2$ nm, $\rho_{Pt} = 31.66$ μΩ·cm, and $\rho_{NiFe} = 78.77$ μΩ·cm.

**C. Power consumption**

Since a relatively high current is required to generate the effective field, it is instructive to estimate the power consumption. In the actual sensor design, two side contacts cover part of the sensor from the two ends and only the central portion is active for sensing. Assuming the active portion has a length of *a/3* and width *b*, the power consumption ($P$) of a single sensor element is given by:

$$P = \frac{1}{3} j_{Pt}^2 b a t_{Pt}^2 \left(1 + \frac{\rho_{Pt}}{\rho_{NiFe}}\right)^2 \frac{\rho_{Pt}}{t_{Pt} + t_{NiFe}\rho_{Pt}/\rho_{NiFe}} \tag{5}$$

where $\rho_{NiFe}$ ($\rho_{Pt}$) is the resistivity of NiFe (Pt) layer, and $j_{Pt}$ is the current density in the Pt layer. Fig.3 shows the calculated power consumption as the function of $j_{Pt}$ using the experimentally obtained resistivity $\rho_{Pt} = 31.66$ μΩ·cm and $\rho_{NiFe} = 78.77$ μΩ·cm. In the calculation, we have set $a/b = 4$, $t_{Pt} = t_{NiFe} = 2$ nm. As can be seen, the power consumption can be controlled within the range of 1 μW to 10 mW by adopting a proper combination of $j_{Pt}$ and $a$. Further power reduction is possible when bilayers with large $H_{FL}/j_{HM}$ are used.

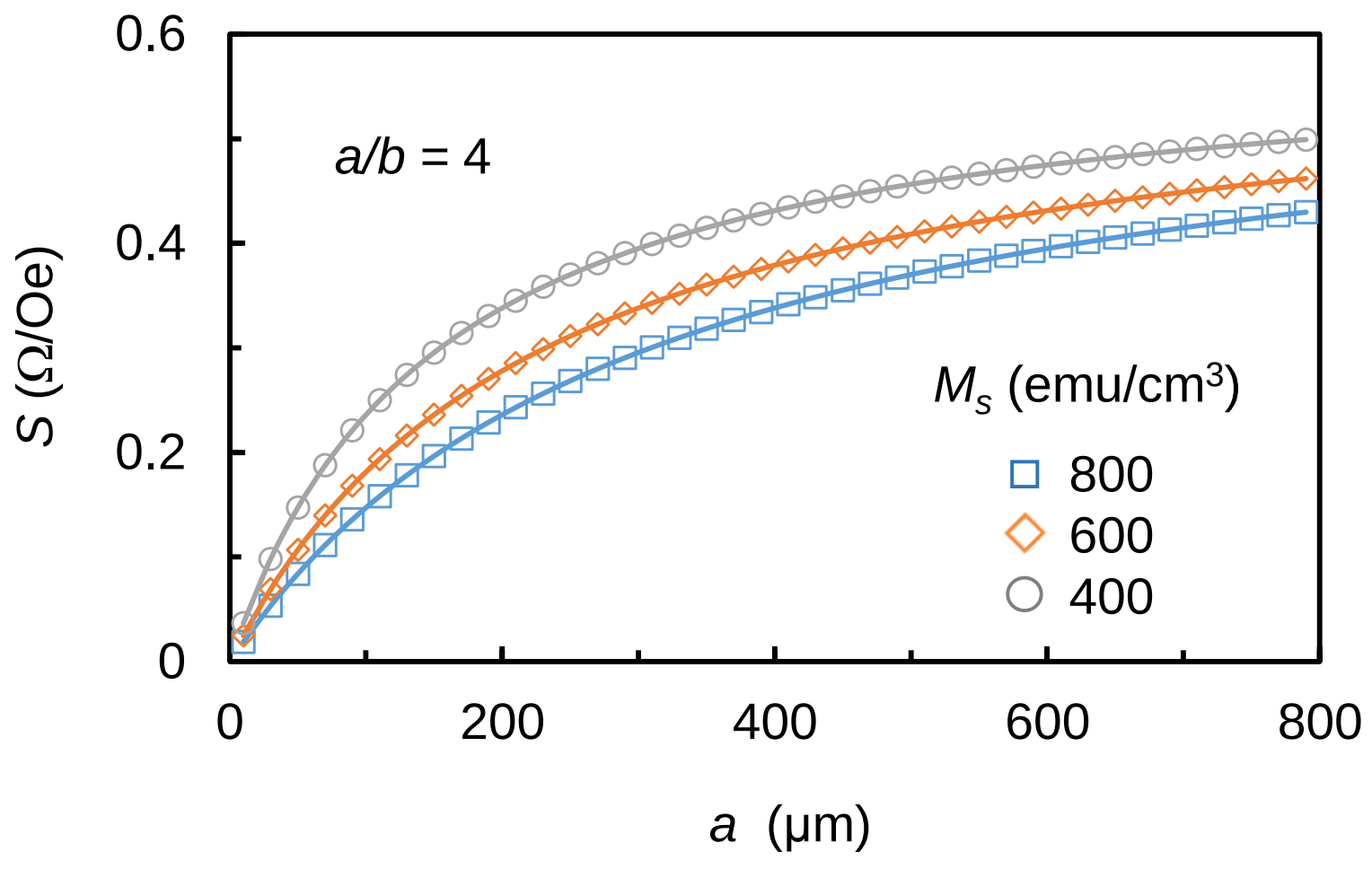

Fig.4. Calculated sensitivity as a function of $a$. Parameters used in the calculation are $L/a$ = 1/3, $a/b$ = 4, $\rho_{Pt}$ = 31.66 μΩ·cm, $\rho_{NiFe}$ = 78.77 μΩ·cm, $\Delta\rho_{NiFe}/\rho_{NiFe}$ =0.07%, $H_k$ = 0.5 Oe, and $t_{NiFe} = t_{Pt} = 2$ nm.

**D. Detection sensitivity**

Besides the power consumption, sensitivity is also essential for magnetic sensing applications. Based on the macro-spin model, we have earlier derived that $\Delta V = \Delta R \alpha I \left(\frac{1}{2} - \frac{\sqrt{2} H_y}{H_k + H_d}\right)$ for a single element sensor, $\alpha < 1$ accounts for the current shunting effect by the HM layer. Therefore, the sensitivity of a single sensor element is given by

$$S_s = \frac{\sqrt{2}a}{3b} \frac{\Delta\rho_{NiFe}}{\rho_{NiFe}} \frac{\rho_{Pt}\rho_{NiFe}}{t_{Pt}\rho_{NiFe} + t_{NiFe}\rho_{Pt}} \frac{1}{H_d + H_k} \tag{6}$$

For a Wheatstone bridge sensor, the sensitivity is $S = 2S_s$. Fig.4 shows the calculated sensitivity as a function of long-axis length $a$. In the calculation, $H_d$ is calculated from $H_d = M_s(N_y - N_x)$. The parameters used are, $L/a$ = 1/3, $a/b$ = 4, $\rho_{Pt}$ = 31.66 μΩ·cm, $\rho_{NiFe}$ = 78.77 μΩ·cm, $\Delta\rho_{NiFe}/\rho_{NiFe} = 0.07\%$, $H_k$ = 0.5 Oe, and $t_{NiFe} = t_{Pt} = 2$ nm. These values are typical for ultrathin NiFe/Pt sensors. The calculation shows that it is possible to achieve a sensitivity in the range of 200 - 400 mΩ/Oe by adjusting the sensor size. As expected, the sensitivity is inversely proportional to the dynamic range, but increases with the sensor size.

**E. Simulated sensor response**

The SOT biasing is ideal for differential sensing using two AMR sensors (as we discussed above, the MR in ultrathin FM/HM bilayers contains both AMR and SMR, but for simplicity we simply call it AMR). As shown schematically in Fig.5a, when the two sensors are oppositely biased by the sensing current, the SOT effective field rotates the magnetization of the two sensors in opposite directions off the easy axis by an angle $\theta$. A linear response with maximum sensitivity is obtained when both magnetizations are $45^o$ away from the easy axis. Although similar magnetic configuration can also be realized using the conventional barber-pole structure, the SOT biasing is much simpler as it does not require patterned metallic strip to direct the sensing current to be $45^o$ from the sensor element's easy axis. This greatly simplifies the fabrication processes for AMR sensors. By assuming $H_d + H_k$ = 1.9 Oe, the sensor's response under the biasing of different $H_{FL}$ can be simulated using the energy minimization method. The simulated AMR curves are shown in Fig.5b (left panel for sensor 1 and right panel for sensor 2) which correspond to $\theta$ = -7.5$^o$ to -75$^o$ for sensor 1 and $\theta$ = 7.5$^o$ to 75$^o$ for sensor 2 with a step size of 7.5$^o$. The corresponding $H_{FL}$ values required are also given in the figure, *i.e.*, -0.2 Oe to -2.66 Oe for sensor 1 and 0.2 Oe to 2.66 Oe for sensor 2. The opposite sign of $H_{FL}$ is a direct result of different current direction in the two sensors. To bias the magnetization into 45° from the easy axis, one only needs a SOT effective field of 1.23 Oe which, as we will discuss shortly in the experimental part, can be readily obtained in NiFe/Pt bilayers with a thin NiFe layer. It is worth noting that at this condition both sensors exhibit maximum sensitivity but they are operating in the different quadrant of the magnetization with respect to the easy axis, which is the key to obtaining linear and symmetrical response from the oppositely biased sensor pair. Fig.5c shows the calculated magnetoresistance as a function of external field $H_y$ at different SOT bias field, *i.e.*, $H_{FL}$ = ±2.66, ±1.72, ±1.23, ±0.82, and ±0.41 Oe. At these $H_{FL}$ values, the corresponding $\theta$ values are $|\theta|$ = 75°, 60°, 45°, 30°, and 15°, respectively. The Oersted field generated by the current was included in the calculation. As can be seen from the figure, a linear response with maximum sensitivity is obtained at $H_{FL}$ = 1.23 Oe, and the sensitivity decreases with either increasing or decreasing the $H_{FL}$ from this value. It is important to note that the linear response is obtained in a very simple bilayer structure without any additional magnetic bias except for the SOT effective field.

The main takeaway of the aforementioned simulation can be summarized as follows. First, the dynamic range of the sensor can be controlled by the shape anisotropy in a large range. The FL effective field is able to bias the magnetization to be 45° away from the easy axis and thereby generating a linear response with a reasonable current density, *i.e.*, $10^6$ – $10^7$ A/cm$^2$. Second, the power consumption for an individual sensor element can be controlled within the range of 0.1 – 10 mW, depending on the sensor size. Third, a sensitivity of over 100 mΩ/Oe can be achieved at a current density around $10^7$ A/cm$^2$. In the simulation, we focused mainly on the current density and sensor size. The effect of aspect ratio and layer thickness on the sensor's performance, as well as the sensor's resolution, have been investigated experimentally and the results will be discussed below.

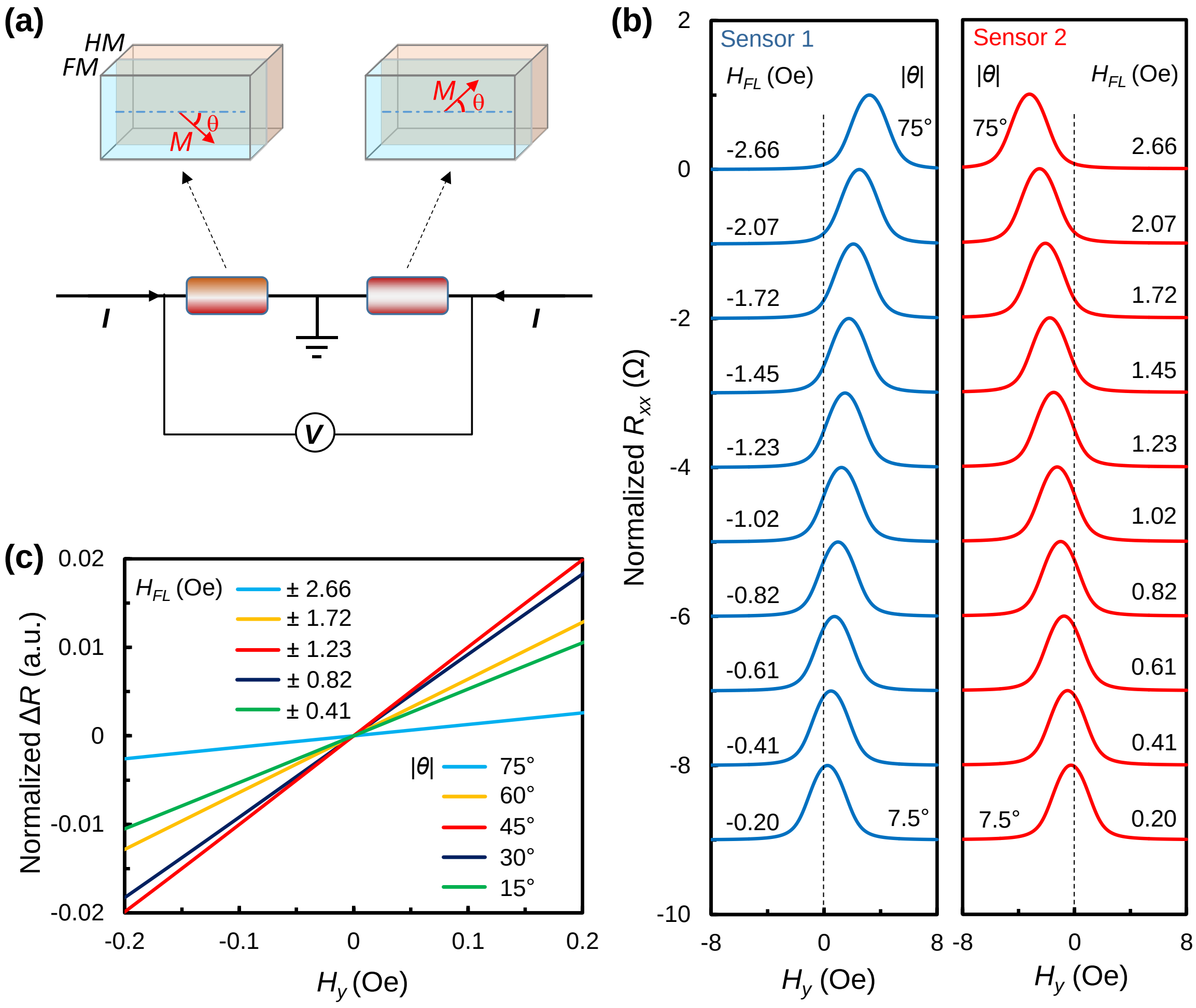

Fig.5. (a) Schematic of a differential AMR sensor with two SOT-biased sensor elements and each sensor element consists of a HM/FM bilayers. (b) Calculated AMR curves for both sensors (left panel: sensor 1 and right panel: sensor 2) at different bias fields: -0.2 Oe to -2.66 Oe for sensor 1 and 0.2 Oe to 2.66 Oe for sensor 2. At these $H_{FL}$ values, $\theta$ = -7.5° to -75° for sensor 1 and $\theta$ = 7.5° to 75° for sensor 2. (c) Calculated magnetoresistance as a function of external field $H_y$ with $H_{FL}$ = ±2.66, ±1.72, ±1.23, ±0.82, and ±0.41 Oe, respectively. We have used $H_k$ = 1.4 Oe and $H_e$ = 0.5 Oe (in the same direction of $H_k$) in the calculation.

## III. EXPERIMETAL

The NiFe/Pt bilayer sensors were fabricated on $SiO_2$/Si substrates with the NiFe layer deposited by e-beam evaporation and Pt by DC magnetron sputtering. Both layers were deposited in a multi-chamber system at a base pressure below $3\times10^{-8}$ Torr without breaking the vacuum. In addition to its role as a HM, the Pt also functions as capping layer to prevent NiFe from oxidation. An in-plane field of ~500 Oe was applied during the deposition to induce a uniaxial anisotropy for the magnetic film. Before patterning into sensor elements, thickness optimization was carried out on coupon films by characterizing its electrical and magnetic properties. From these measurements, basic properties such as magnetization and resistivity were obtained. Magnetic measurements were carried out using a Quantum Design vibrating sample magnetometer (VSM) with the samples cut into a size of 3 mm × 2.5 mm. The resolution of the system is better than $6\times10^{-7}$ emu. All electrical measurements were carried out at room temperature.

## IV. EXERIMENTAL RESULTS AND DISCUSSION

### A. FL effective field in NiFe/Pt bilayers

In order to quantify the $H_{FL}/j_{Pt}$ ratio experimentally, we measured the $H_{bias}$ ,which is the sum of $H_{FL}$ and $H_{Oe}$ in $y$-direction, as a function of current density for NiFe($t_{NiFe}$)/Pt(2) bilayer structures with $t_{NiFe}$ = 1.8, 2, 3 and 4 nm by using the 2nd order planar Hall effect (PHE) method.[18,21] Here, the numbers inside the brackets denote thickness in nm. Details of measurement procedure can be found in our previous work.[22,23] The devices used for extracting $H_{bias}$ are fabricated directly on $SiO_2$/Si substrates without any seed layer using combined technique of sputtering/evaporation and lift-off. The devices are ellipsoid shaped with a long axis of 3000 μm and short axis of 375 μm while an easy axis is induced in the long axis (or $x$-) direction by applying an external in-plane magnetic field during deposition. As summarized in Fig.6a, the $H_{bias}$ values extracted directly from experiments scale linearly with the current density in Pt layer at different NiFe thickness. The ratio $H_{FL}/j_{Pt}$ is obtained and shown in Fig. 6b as a function of the NiFe layer thickness after subtracting the contribution from the Oersted field in the bilayers. For NiFe($t_{NiFe}$)/ Pt($t_{Pt}$) bilayer with a lateral dimension of $a \times b$, where *a (b)* is the long axis length (short axis length) of the sensor element, the Oersted field in the middle of NiFe layer due to current in the Pt layer is given by $\frac{H_{Oe}}{j_{Pt}} = \frac{t_{Pt}}{2}$ when $b >> t_{Pt}, t_{NiFe}$. In the present case, $t_{Pt}$ = 2 nm, therefore $H_{Oe}/j_{Pt}$ = 0.126 Oe/($10^6$A/cm$^2$). The estimated $H_{Oe}/j_{Pt}$ ratio is shown in Fig. 6b in dotted-line, which alone is apparently too small to account for the experimentally observed biasing field ($H_{bias}$) in *y*-direction and also the NiFe thickness dependence of $H_{bias}/j_{Pt}$. As we mentioned in Section II, the SOT is dependent on the spin mixing conductance at the interface, which varies from sample to sample. It is, therefore, more meaningful to focus on the NiFe thickness dependence rather than absolute values of $H_{FL}/j_{Pt}$. As such, we may express the SOT efficiency as $\frac{H_{FL}}{j_{Pt}} = \frac{\hbar}{2e}\frac{\theta_{SH}\alpha}{\mu_0 M_s t_{NiFe}}$, where $\alpha$ is a parameter that contains spin mixing conductance at NiFe/Pt interface, thickness and spin diffusion length of Pt but is independent of NiFe thickness, $t_{NiFe}$. $\theta_{SH}\alpha$ is equivalent to the effective spin Hall angle introduced by Nan *et al.*[16] As the saturation magnetization at small thickness is usually different from its bulk value, we measured the saturation magnetization of NiFe($t_{NiFe}$)/Pt(2) bilayers at different NiFe thicknesses using a vibrating sample magnetometer, and the $M_S$ values obtained are $\mu_0 M_s$ = 0.65, 0.74, 0.97 and 1.02 T for $t_{NiFe}$ = 1.8, 2, 3 and 4 nm, respectively. Using these values, the experimental data shown in Fig. 6b can be fitted reasonably well by assuming $\theta_{SH}$ = 0.15 (ref. [24]) and $\alpha$ = 0.122 (note $H_{FL} = H_{bias} - H_{Oe}$). These results confirm that the main contribution of the experimentally observed biasing effective field is from the SOT effective field. In addition to its much larger strength as compared to $H_{Oe}$, the $H_{FL}$ is also more uniform in samples with a finite size, especially at the edge of the samples. It should be pointed out that even at this proof-of-concept stage, as shown in Fig. 6(b), the largest $H_{FL}/j_{Pt}$ ratio achieved so far is 0.76 Oe/($10^6$ A/cm$^2$) at a NiFe thickness of 1.8 nm. To obtain a SOT effective field of 1 Oe, we only need a current density of about 1.3 × $10^6$ A/$cm^2$, which is considered moderate even for conventional AMR sensor with additional transverse biasing.

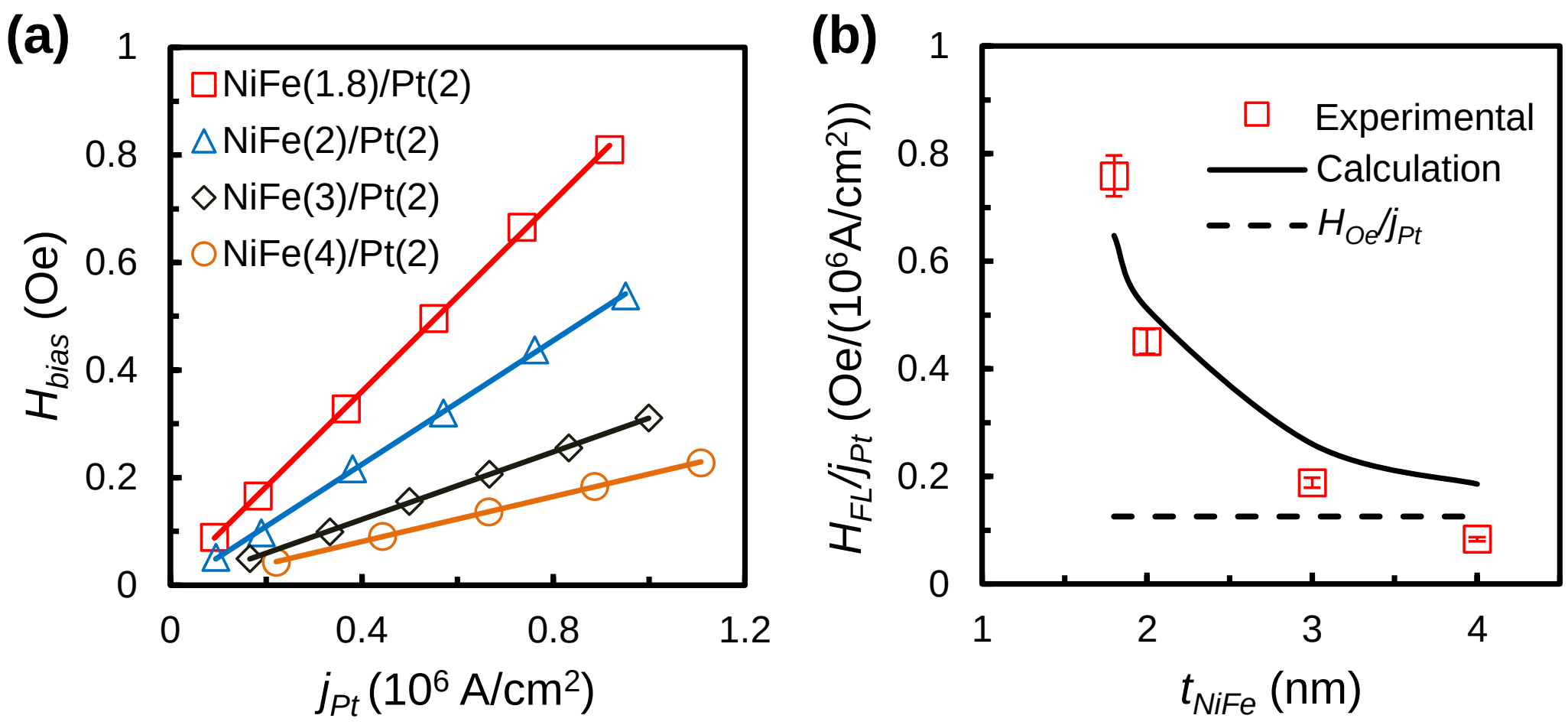

Fig.6. (a) Experimentally determined $H_{FL}$ as a function of $j_{Pt}$ for NiFe($t_{NiFe}$)/Pt(2) bilayers with $t_{NiFe}$ = 1.8, 2, 3 and 4 nm, using 2nd order PHE measurement. (b) Extracted $H_{FL}/j_{Pt}$ ratio as a function of $t_{NiFe}$ for NiFe($t_{NiFe}$)/Pt(2) (square symbol) and calculated Oersted field ($H_{Oe}$) at the center of NiFe layer (dotted-line). Solid-line is the fitting using Eq. (3) assuming $\theta_{SH}$ = 0.15 and $\alpha$ = 0.122.

### B. Linearization by SOT effective field

To verify the SOT-biasing effect and compare it with the simulation results shown in Fig.5, we fabricated two ellipsoid shaped NiFe(2)/Pt(2) sensors with a long axis length of 1500 μm and an aspect ratio of 4:1. As shown in the scanning electron micrograph (SEM) in Fig. 7a, the two sensors are connected at the middle and form a Wheatstone bridge with two external resistors $R_1$ and $R_2$. The values of $R_1$ and $R_2$ are adjusted slightly to account for the process induced small difference in the resistance of the two sensors. When a current source is connected to the bridge as depicted in the Fig. 7a, the magnetization of the two sensors are rotated to opposite directions with respect to the easy axis, leading to a linear response to the external field which is detected as a voltage signal from the other two terminals of the bridge. Fig.7b shows the MR curves of both sensors (left panel: sensor 1 and right panel: sensor 2) at bias current densities ranging from 1.9 × $10^5$ A/cm$^2$ to 1.9 × $10^6$ A/cm$^2$. When the bias current increases, the curves are shifted in opposite directions. A nearly linear region with maximum sensitivity is obtained for both sensors near $H_y$ = 0 when the current density is around 1.9 × $10^6$ A/cm$^2$. As with the simulated curves in Fig. 5b, the two sensors are operating in the different quadrant of the magnetization with respect to the easy axis, leading to linear and symmetrical responses when connected in a bridge in Fig. 7a. Shown in Fig. 7c are the output signals as a function of $H_y$ at different bias current densities. The output signal ΔR is defined as the output bridge voltage divided by the current passing through the bilayer sensor element. The sensor exhibits good linearity with a maximum sensitivity at $j_{Pt} = 1.9 \times 10^6$ A/cm$^2$, which decreases by increasing or reducing the bias current. This is in good agreement with the simulation results shown in Fig. 5c. The results demonstrate clearly good tunability of SOT-biasing.

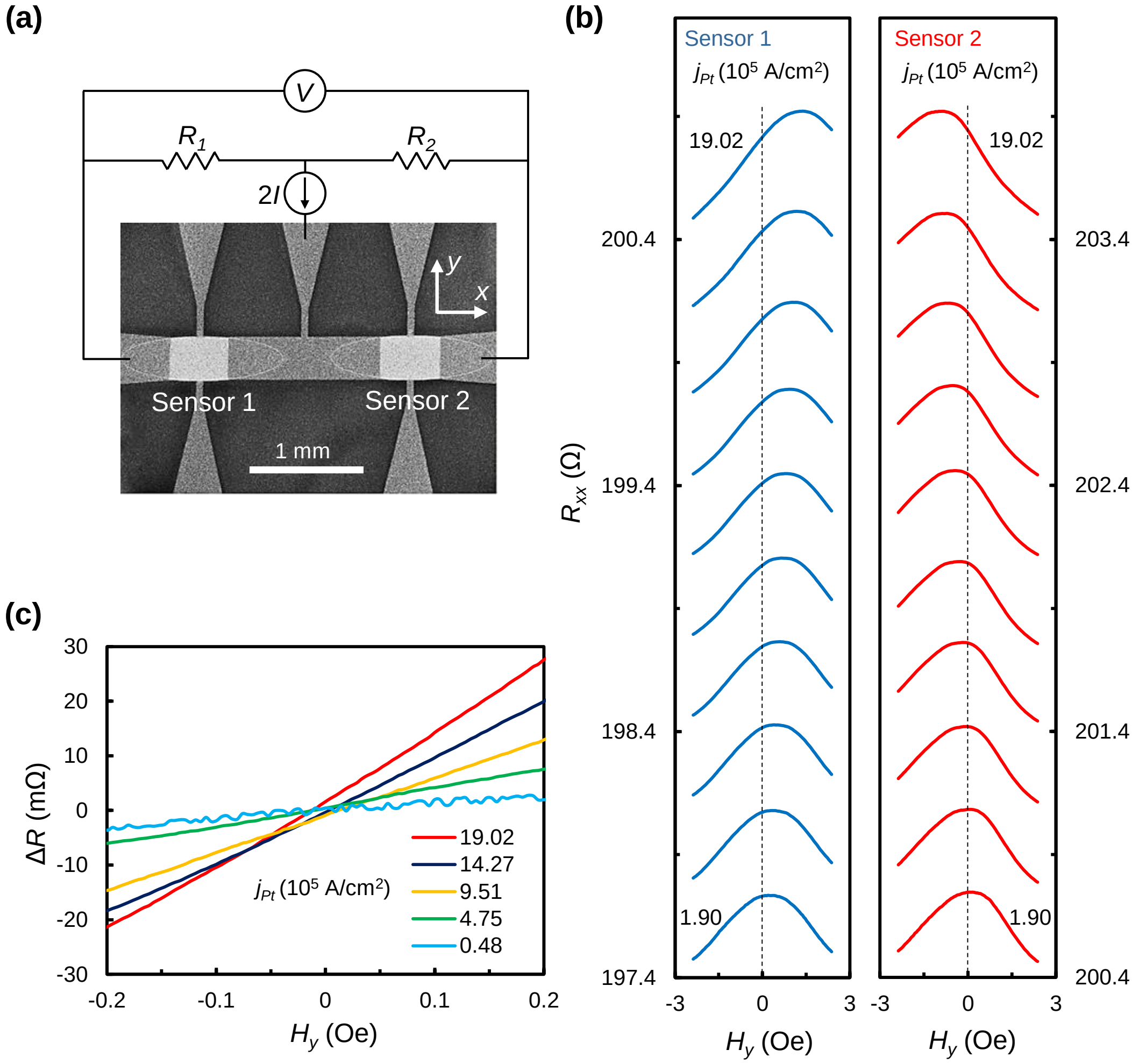

Fig.7. (a) Scanning electron micrograph and schematic of AMR bridge sensor. Scale bar: 1 mm. (b) Measured MR curves of both sensors (left panel: sensor 1 and right panel: sensor 2) at different bias current densities: 1.9 × $10^5$ - 1.9 × $10^6$ A/cm$^2$. (c) Output signals as a function of $H_y$ at different bias current densities.

### C. SOT-biased Wheatstone bridge sensors

In order to evaluate the field sensing performance of SOT biased sensors with different dimensions, we fabricated full Wheatstone bridge sensors with ellipsoidal shape in NiFe(1.8)/Pt(2) bilayers. The long to short axis ratio is fixed at $a/b = 4$, with $a = 800$, 400 and 200 μm, respectively. The distance ($L$) between the two electrical contacts for each sensor element is kept $a/3$. In order to minimize the influence of earth field, both the sensors and Helmholtz coils for generating the field were placed inside a magnetically shielded cylinder made of 7 layers of μ-metals. Fig.8a shows the scanning electron micrograph of the four sensor elements with $a = 800$ μm, which are connected to form a Wheatstone full bridge. When a current source is connected to the top and bottom terminals of the bridge sensor as depicted in the Fig. 8a, the magnetization of the sensor elements, 1 and 4, are rotated to the direction opposite to that of the sensor elements, 2 and 3, with respect to the easy axis, leading to a linear response to the external field which is detected as a voltage signal from the other two terminals of the bridge. Fig. 8b shows the AMR curves of all the four sensor elements at the same bias current densities of $3.67 \times 10^5$ A/cm$^2$ at which a nearly linear response region with maximum sensitivity is achieved at zero external field. The MR response curves of the four sensing elements all exhibit a field shift that is proportional to the bias current with the shift direction determined by the polarity of the current. This is consistent with the SOT biasing scheme as discussed in Section II, and it clearly rules out Joule heating as the cause for the rotation of magnetization. Shown in Fig. 8c are the output signals as a function of $H_y$ at different bias current densities. The sensor exhibits good linearity with a maximum sensitivity at $j_{Pt} = 3.67 \times 10^5$ A/cm$^2$, which decreases by increasing or reducing the bias current. From the slope of the response curve in Fig. 8c, we can extract the maximum sensitivity of the sensor which is 0.548 Ω/Oe. This is much larger than the simulated sensitivity in Fig.4 at the same current density. This is because in the simulation, the parameters used are for typical NiFe/Pt heterostructures, which are not necessary the same for all the sensors. In particular, in the present case, the NiFe is only 1.8 nm, which leads to a much better SOT efficiency.

In order to examine the detection limit of these SOT-biased full bridge AMR sensors, we performed AC field sensing experiments and analyzed the waveform of the output signal. During these experiments, an AC magnetic field with various magnitudes and fixed frequency of 0.1 Hz was applied in *y*-direction, while the sensor output was recorded with respective to time. The output signals of the sensor with $a$ = 800 μm, when being biased at a current density of $j_{Pt} = 3.67 \times 10^5$ A/cm$^2$ and used to detect a 0.1 Hz AC field with amplitudes ranging from 10 nT to 30 μT are summarized in Fig.8d. The amplitude of output signal decreases with the amplitude of applied AC external field, and is eventually masked out by the noise. To have a clearer view of the background noise, we re-display the output signals obtained at AC field amplitudes of 30 μT, 500 nT and 10 nT in Fig.8e. The corresponding Fourier transform (FT) of the output waveforms is shown in Fig. 8f. As can be seen, a clear peak at 0.1 Hz can be identified for all three cases. However, as the amplitude of the applied AC field decreases further to below 10 nT, the 0.1 Hz peak becomes indistinguishable (not shown here).

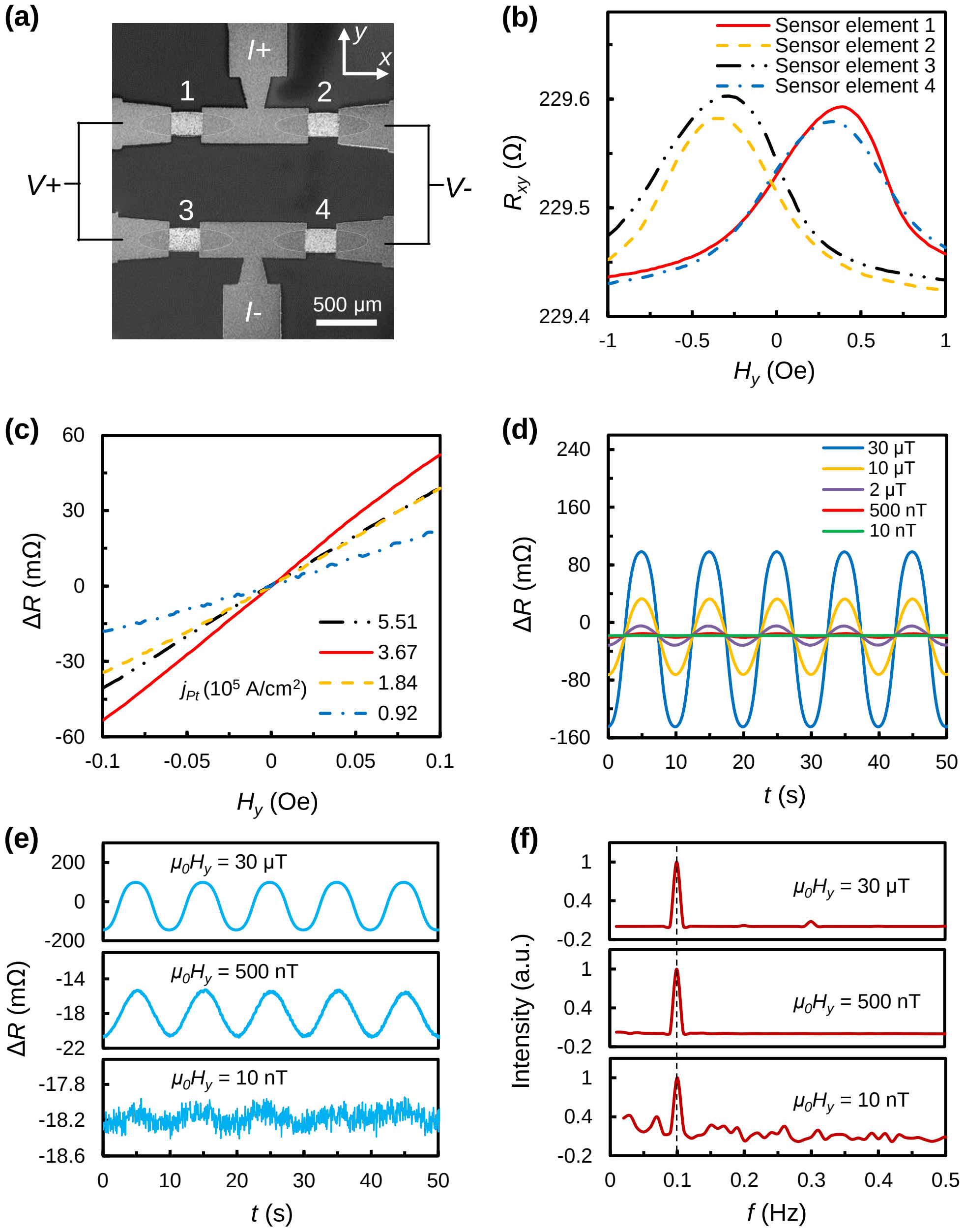

Fig.8. (a) SEM image and schematic of the SOT biased Wheatstone bridge sensor. Scale bar: 500 μm. (b) Measured AMR curves of all the four sensor elements at the same bias current densities of 3.67 × $10^5$ A/cm$^2$. (c) Output signals as a function of $H_y$ at different bias current densities. (d) AMR response of the bridge sensor to an AC magnetic field with a frequency of 0.1 Hz but with a varying amplitude: 10 nT, 500 nT, 2 μT, 10 μT and 30 μT. The AC field is in the hard axis direction. The sensor is biased at $j_{Pt}$ = 3.67 × $10^5$ A/cm$^2$. (e) Output signals extracted from (d) with different amplitude: 30 μT, 500 nT and 10 nT. (f) Fourier transform of the waveforms in (e).

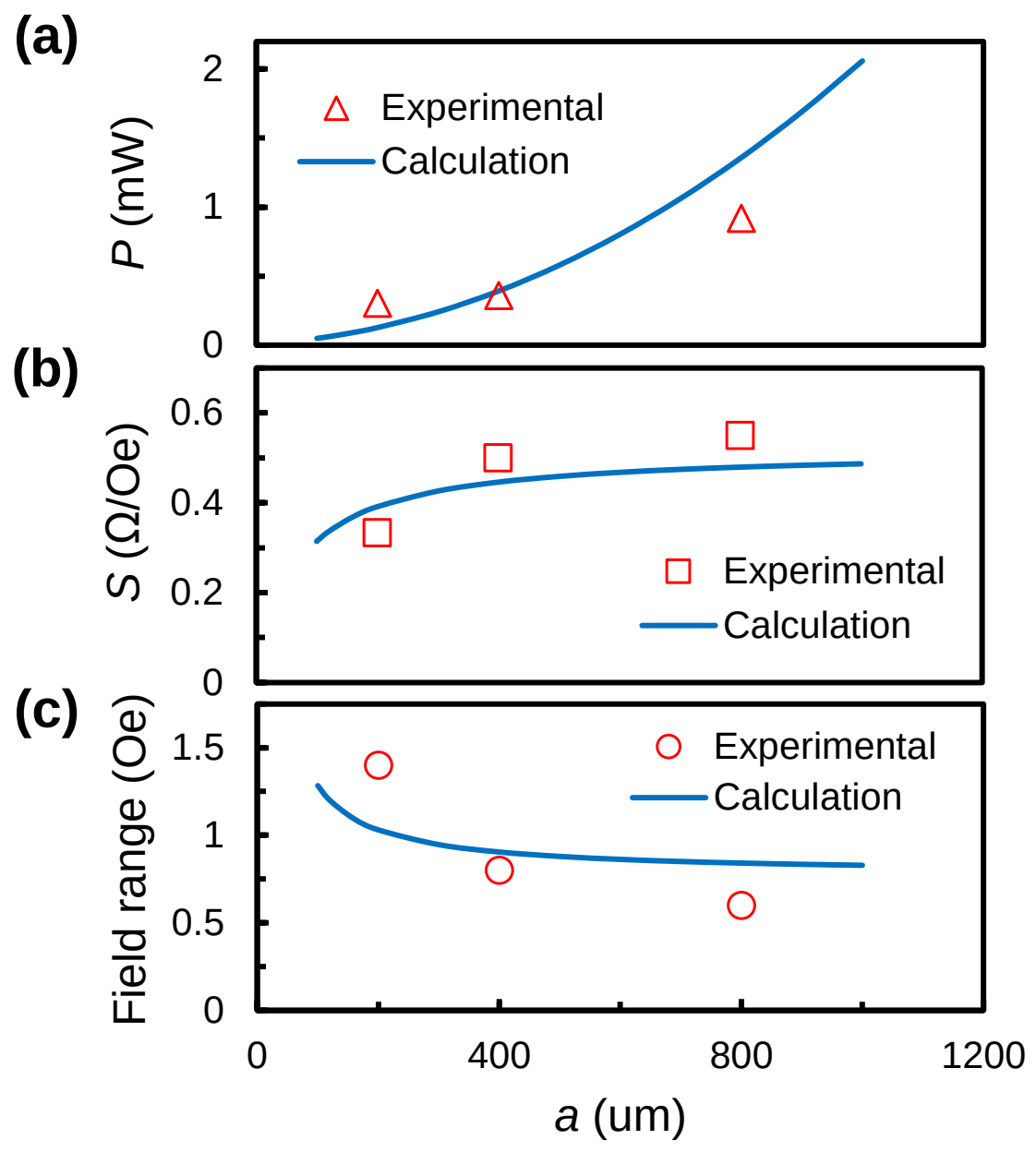

Fig.9. Dependence of power consumption (a), sensitivity (b), and dynamic range (c) on the long axis length *a* (symbols: experiment; solid curve: simulation).

Similar measurements were performed on the other two sensors with $a$ = 400 and 200 μm, respectively. The bias current densities required to achieve linear response with maximum sensitivity at zero external field are $4.59 \times 10^5$ A $cm^{-2}$ and $8.44 \times 10^5$ A $cm^{-2}$, for $a$ = 400 and 200 μm, respectively. In AC field sensing measurements, the resolution of the two sensors turned out to be 20 and 70 nT, for the sensors with $a$ = 400 and 200 μm, respectively. In Fig.9, we show the sensor size dependence of power consumption, sensitivity, and dynamic range of the three sensors (symbols). By changing the long and short axis length (*a* and *b*), the shape anisotropy of the sensor can be changed accordingly; this leads to tunable bias current density, power consumption, sensitivity and dynamic range. The solid curves in Fig.9 are calculated based on the formulas derived in previous section, using the parameters *L/a* = 1/3, a/b = 4, $\rho_{Pt}$ = 31.66 μΩ·cm, $\rho_{NiFe}$ = 78.77 μΩ·cm, $\mu_0 M_s$ = 0.65 T, $\Delta\rho_{NIFe}/\rho_{NiFe}$ = 0.06%, $H_k$ = 0.5 Oe, $t_{NiFe}$ = 1.8 nm and $t_{Pt}$ = 2 nm. From the calculation results, we can observe, by reducing *a* from 1000 μm to 100 μm, the power consumption decreases significantly from 2.06 mW to 0.05 mW, the sensitivity decreases from 243.2 to 157.0 mΩ/Oe and the dynamic range increases from 0.83 to 1.28 Oe. These changes are attributed to the increased shape anisotropy and reduced current as the dimension decreases. The agreement between experimental and simulated results shows clearly that it is possible to tune the sensor's power consumption, sensitivity and dynamic range via adjusting the dimensions.

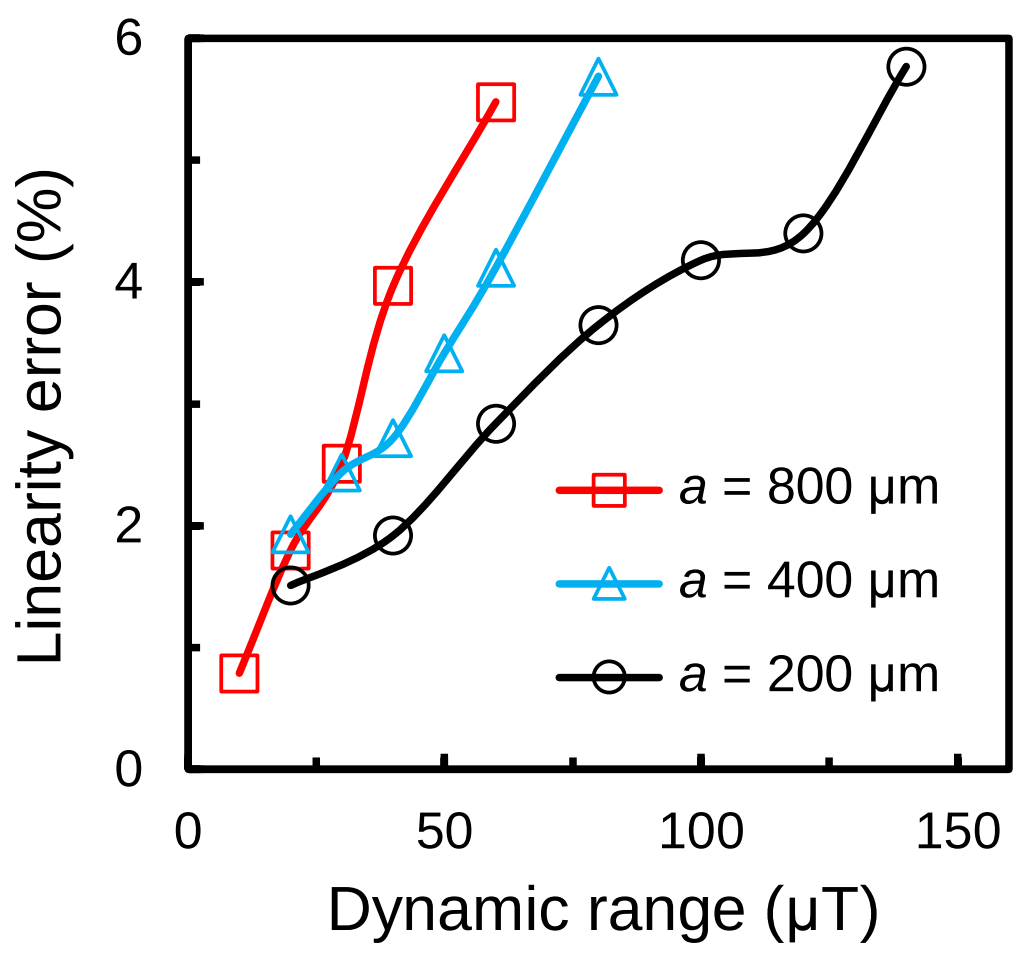

Fig.10. Linearity error versus dynamic range for NiFe(1.8)/Pt(2) ellipsoidal sensors with a = 800, 400 and 200 μm, respectively.

All the sensors exhibit good linearity at low field, but the linearity error increases with the applied field. Fig.10 shows the experimentally extracted linearity error as a function of the dynamic range. Here, the linearity error (%) is defined as the deviation of the sensor output curve from a specified straight line over a desired dynamic range. It is clearly shown that the linearity error increases as the dynamic range increases, which is typical for AMR sensors. Compared to commercial AMR sensors, the dynamic range of SOT-biased NiFe/Pt bridge sensor is small, mainly because of the relatively small $H_{FL}$ (~1 Oe) in this specific material system. Hence, in order to have good linearity, the external field must be smaller than the transverse biasing field which in this case is the sum of $H_{FL}$ and $H_{Oe}$. By defining the working field range as the dynamic range that gives a linearity error below 6%, the field ranges for the three sensors are obtained and found to correlate well with calculated values, as shown in Fig.9c. It is important to note that the sensitivity is inversely proportional to the dynamic field range, the dimension of the sensor has to be optimized in order to achieve desired performance.

## D. Noise characterization on SOT-biased Wheatstone full bridge sensor

In order to estimate the detectivity of the sensor, standard noise characterization was performed using the setup as shown in Fig. 11(a). The noise measurements were carried out in a magnetically shielded cylinder made of 7 layers of μ-metals. The sensor was powered by a battery and its bridge output voltage was amplified by a low-noise amplifier (DLPVA-100-B). Dynamic signal analyzer (Agilent 35670A) was used to acquire the noise power spectrum.

The equivalent field noise, or detectivity, can be obtained by dividing the noise voltage spectral density over the sensor sensitivity. The detectivity for the NiFe(1.8 nm)/Pt(2 nm) Wheatstone full bridge SOT biased sensor with a dimension of 800 μm × 200 μm is shown in Fig.11(b). The result shows that the sensor has a detectivity of about 2.8 nT/√Hz at 1 Hz and is able to detect sub-nano Tesla field above 10 Hz. Upon further process optimization, we believe the detectivity can be improved further.

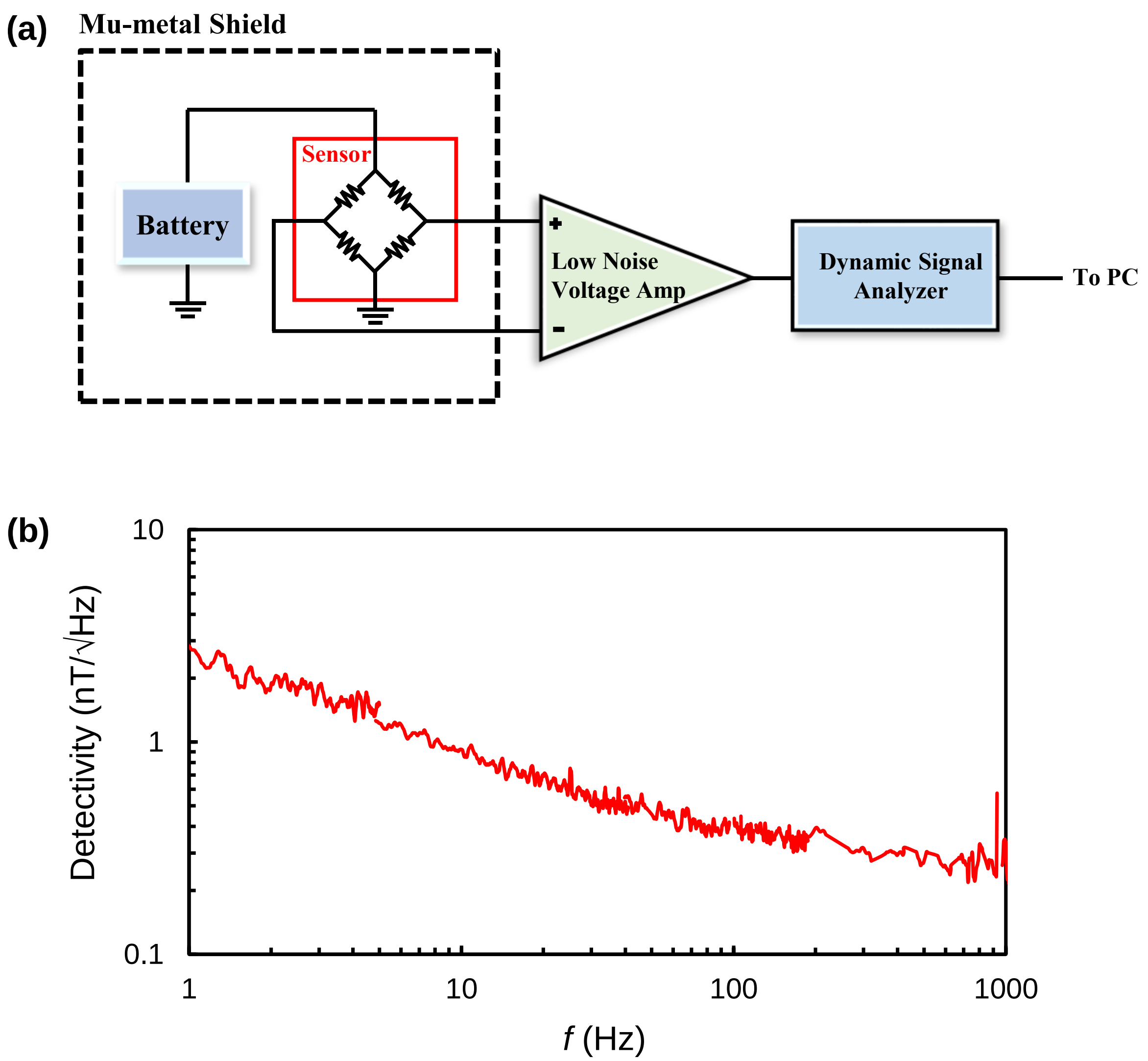

Fig.11. (a) Block diagram of the noise measurement system. (b) Detectivity of NiFe(1.8 nm)/Pt(2 nm) Wheatstone full bridge SOT biased sensor with a dimension of 800 μm × 200 μm.

### E. On-chip current detection using SOT-biased AMR sensor

Given its simple structure and ultrathin thickness, SOT-biased sensors can be potentially used in on-chip monitoring of electric current. As a proof-of-concept experiment, we fabricated a Wheatstone bridge sensor with four ellipsoidal shape sensing elements comprised of NiFe(1.8)/Pt(2) bilayers; the entire sensor is then covered with a 200 nm $SiO_2$ isolation layer, followed by a copper layer with thickness (width) of 500 nm (2000 μm), as shown schematically in Fig. 12a and b. The dimensions of the sensing elements are kept the same as those shown in Fig. 8a. Current sensing measurements were carried out in a magnetically shielded cylinder with 7 layers of μ-metal sheets. We first established the linear operation region of the sensor by subjecting the sensor to the stray field generated by the current in the copper wire. Shown in Fig. 12c are the bridge output signals as a function of current in the copper wire at different bias current densities in the Pt layer of the sensor elements. The sensor exhibits good linearity with a maximum sensitivity of 1.54 Ω/A at $j_{Pt} = 1.1 \times 10^6$ A/cm$^2$; the sensitivity decreases with either the increase or decrease of the bias current. The current density required to achieve maximum sensitivity is slightly higher than that for the sensor with the same structure shown in Fig. 8a. This is may be caused by the overlaid $SiO_2$ layer on the sensor elements; further study is required to optimize the deposition processes.

In order to correlate the current generated stray field with external field, we performed field sweeping measurement on the same sensor using Helmholtz coils, and a maximum sensitivity of 487.2 mΩ/Oe is obtained. This gives a field to current ratio of 3.16 Oe/A, corresponding to a field to current density ratio of 3.16 Oe/(10$^5$ A/cm$^2$), for the copper wire at the sensor plane. To compare with the measurement results, we calculated the Oersted field generated by the copper wire using three-dimensional finite element analysis. In order to shorten the calculation time, the dimension of the copper wire was scaled down to 8 μm with the thickness unchanged. The current densities used for the calculation were kept the same as those of the actual device when a current of 0 - 1 A flows in the copper wire. Fig.12d shows the calculated Oersted field ($H_y$) at the sensor plane as a function of the current density in the Cu layer (inset shows the distribution of *y*-component of the Oersted field in *yz* plane at $j_{Cu} = 1 \times 10^5$ A/cm$^2$). The slope of the curve is 3.141 Oe/($10^5$ A/cm$^2$), which is close to the experimentally extract value of 3.16 Oe/($10^5$ A/cm$^2$). By defining the working range as the dynamic range that gives a linearity error below 6%, the working current range for the this sensor is about 0.3 A. Similar AC field sensing experiments have been performed, in order to obtain the detection limit of the sensor. To this end, an AC current with varying magnitude but fixed frequency of 0.1 Hz was applied to the Cu wire, and the output of the sensor was recorded for a certain period of time. This sensor was biased at a current density of $j_{Pt} = 1.1 \times 10^6$ A/cm$^2$. The output signals for current with amplitude of 100 mA, 2 mA and 100 μA are shown in Fig.12e, respectively. The corresponding Fourier transforms of the output waveforms are shown in Fig. 12f. A peak at 0.1 Hz can be clearly identified for all the three current amplitudes. However, as the amplitude of the AC current decreases further to below 100 μA, the 0.1 Hz peak becomes hardly observable (not shown here). Therefore, the current detection resolution of this specific sensor is around 100 μA.

Before concluding, we comment briefly on thermal drift which is an important issue to be addressed for any practical applications of magnetic field sensors. As we mentioned in Section IIA, the MR of the sensor reported in the present work is dominated by SMR. In the case of NiFe/Pt bilayers, we found that the SMR is relatively insensitive to temperature as compared to AMR. This in combination with small driving current help to reduce the thermal drift. A systematic study of thermal drift is currently on-going and the results will be reported elsewhere.

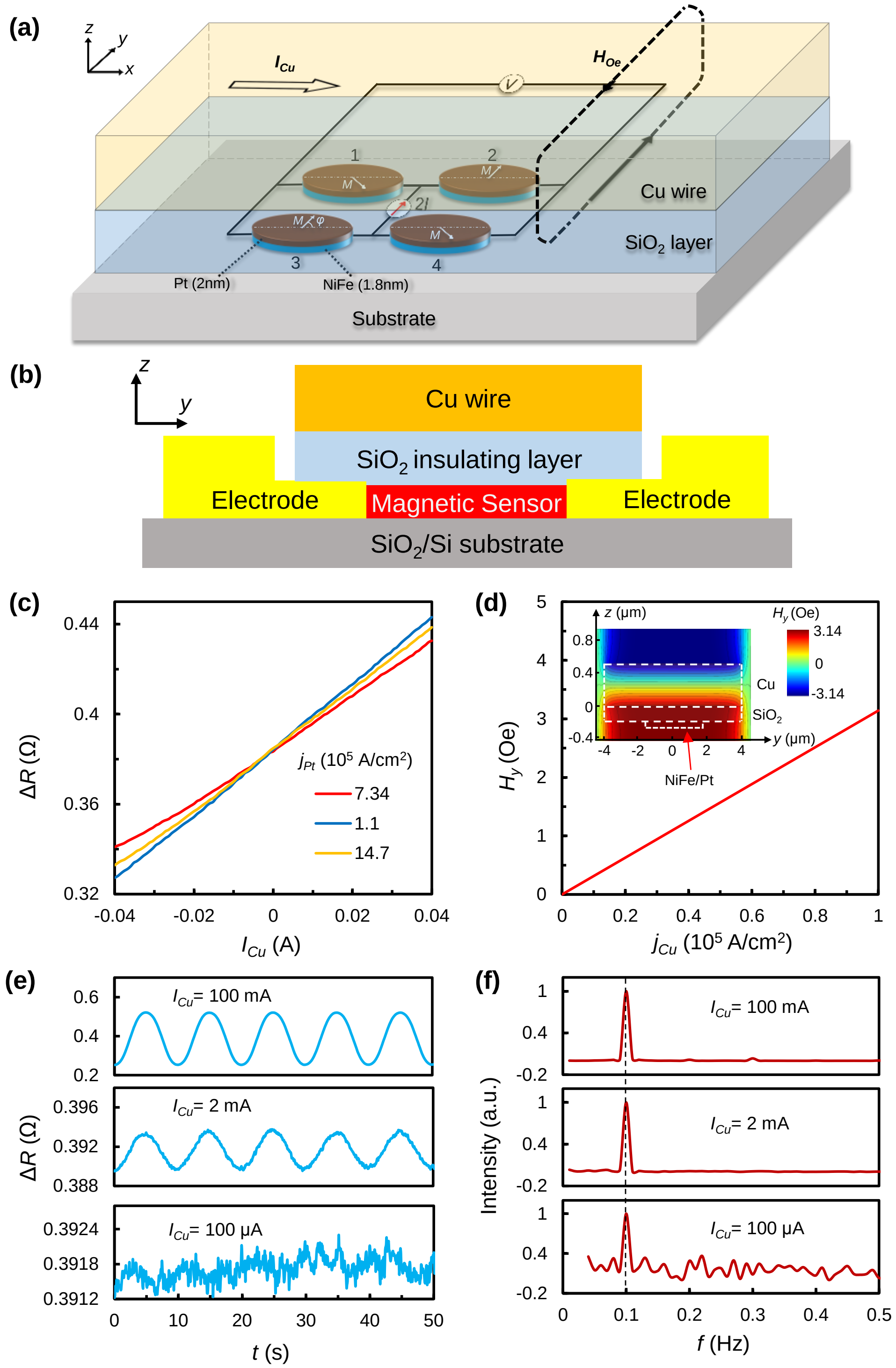

Fig.12. (a) Schematic of the SOT-biased sensor device for on-chip current detection based on NiFe(1.8)/Pt(2) bilayer structure. (b) Schematic cross-section of the device in (a). (c) Output signals as a function of current in the on-chip copper wire at different $j_{Pt}$. (d) Calculated Oersted field ($H_y$) at the sensor plane as a function of current density in the Cu wire (inset: distribution of $y$-component of the Oersted field in $yz$ plane at $j_{Cu}$ = 1 × $10^5$ A/cm$^2$). (e) Output waveforms corresponding to AC current with varying amplitude: 100 mA, 2 mA and 100 μA, but a fixed frequency of 0.1 Hz. The sensor is biased at $j_{Pt}$ = 1.1 × $10^6$ A/cm$^2$. (f) Fourier transform of the waveforms in (e).

## V. CONCLUSIONS

We have performed a systematic study of SOT-biased AMR sensors by focusing on the dimension and current biasing effect. By varying the Pt current density in the range of $j_{Pt} = 1 \times 10^5 - 10^7 A/cm^2$, the sensor dynamic range can be varied in the range of 0.1 - 10 Oe through manipulation of the shape anisotropy. With an aspect ratio of 4:1, the power consumption of a single sensor can be controlled in the range of 1μW – 1mW by varying the sensor size. Under typical conditions, a sensitivity in the range of 0.1-0.5 Ω/Oe can be obtained, depending on the size of the sensor. Experimental results of sensors with selected sizes agree well with simulated results. A field resolution of 10 nT is obtained for a sensor with an aspect ratio of 4:1, size $a$ = 800 μm, and Pt current density of $j_{Pt}$ = 3.67 × $10^5$ A/cm$^2$. As a proof-of-concept experiment, we demonstrated on-chip current detection using the sensor with current resolution of 100 μA in the specific sensor design. This work opens new opportunity for further exploitation of the SOT technology in sensor applications.

## ACKNOWLEDGMENTS

The authors wish to thank Ying Wang for his help with the SEM observation. Y.H.W. would like to acknowledge support by the Singapore National Research Foundation, Prime Minister's Office, under its Competitive Research Programme (Grant No. NRF-CRP10-2012-03) and National University of Singapore under its Academic Research Fund (AcRF) Tier 1 (Grant No. R-263-000-A95-112). Y.H.W. is a member of the Singapore Spintronics Consortium (SG-SPIN).